\documentclass[prd,preprint,showpacs]{revtex4}
\usepackage{amsmath}
\usepackage{latexsym}
\usepackage{amsfonts}
\usepackage{graphicx}
\begin{document}

\title{Latest Observational Constraints on Cardassian Models}
\author{Tai-Shan Wang and Puxun Wu }

\affiliation{Department of Physics \& Center for Astrophysics,
Tsinghua University, Beijing 100084, China}

\begin{abstract}
Constraints on  the original Cardassian model and the modified
polytropic Cardassian model are examined from the latest derived 397
Type Ia supernova (SNe Ia) data, the size of baryonic acoustic
oscillation peak from the Sloan Digital Sky Survey (SDSS), the
position of first acoustic peak of the Cosmic Microwave Background
radiation (CMB) from the five years Wilkinson Microwave Anisotropy
Probe (WMAP), the x-ray gas mass fractions in clusters of galaxies,
and the observational H(z) data. In the original Cardassian model
with these combined data set, we find
$\Omega_{m0}=0.271^{+0.014}_{-0.014}, n=0.035^{+0.049}_{-0.049}$ at
$1 \sigma$ confidence level. And in the modified polytropic
Cardassian model, we find that
$\Omega_{m0}=0.271^{+0.014}_{-0.015}$, $n=-0.091^{+0.331}_{-1.908}$
and $\beta=0.824^{+0.750}_{-0.622}$ within $1\sigma$ confidence
level. According to these observations, the acceleration of the
universe begins at $z_T=0.55^{+0.05}_{-0.05} (1\sigma)$ for the
original Cardassian model, and at $z_T=0.58^{+0.12}_{-0.12}
(1\sigma)$ for the modified polytropic Cardassian model. Evolution
of the effective equation of state $w_{eff}$ for the  modified
polytropic Cardassian model is also examined here and results show
that an evolutionary quintessence dark energy model is favored.
\end{abstract}

\pacs{98.80.Es, 95.35.+d, 98.80.Jk}  \maketitle

\section{Introduction}
The astrophysical observations of recent years, including Type Ia
supernovae (SNe Ia;
~\cite{bie98,bie04,per99,ton03,kno03,bar04,kow08}),  the large scale
structure   \cite{eis05}, and the cosmic microwave background
radiation (CMB; \cite{bal00,ber00,jaf01,spe03,spe07,komtemp}) et al,
show that the present expansion of our universe is accelerating.  In
order to explain this observed accelerating expansion, a large
number of cosmological models have been proposed by cosmologists.
There are two main categories of proposals. The first ones (dark
energy models) are proposed by assuming the existence of an energy
component with negative pressure in the universe, this dark energy
dominates the total energy density of the universe and drives its
acceleration of expansion at late times. Currently there are many
candidates for dark energy, such as the cosmological constant with
equation of state $\omega_{DE} = p_{DE}/\rho_{DE} = -1$ where
$p_{DE}$ and $\rho_{DE}$ are pressure and density of the dark
energy, respectively \cite{car92}, the quiessence whose equation of
state $\omega_{Q}$ is a constant between $-1$ and $-1/3$
\cite{ala03}, and the quintessence which is described in terms of a
scalar field $\phi$ \cite{cal98,sah00}. The other proposals suggest
that general relativity fails in the present cosmic scale, and the
extra geometric effect is responsible for the acceleration, such as
the braneworld models which explain the acceleration through the
fact that the general relativity is formulated in 5 dimensions
instead of the usual 4 \cite{csa00}, and the Cardassian models which
investigate the acceleration of the universe by a modification to
the Friedmann-Robertson-Walker (FRW) equation \cite{fre02}.

In this work we focus on the Cardassian models, including the
original Cardassian model and the modified polytropic Cardassian
model. The original Cardassian model is based on the modified
Friedmann equation and has two parameters $\Omega_{m0}$ and $n$.
The modified polytropic Cardassian model can be obtained by
introducing an additional parameter $\beta$ into the original
Cardassian model which reduces to the original model if $\beta=1$.

As we know, many observational constraints have been placed on
Cardassian models, including those from the angular size of high-z
compact radio sources \cite{zhu02}, the SNe Ia
\cite{wan03,dav07,zhu03,cao03,szy04,god04,fri04,zhu04,ben05,ama05,ben06},
the shift parameter of the CMB\cite{dav07,fri04,ama05,yi07}, the
baryon acoustic peak from the SDSS \cite{dav07,yi07}, the
gravitational lensing \cite{alc05}, the x-ray gas mass fraction of
clusters \cite{zhu04,zhu042}, the large scale structure
\cite{ama05,mul03,fay06}, and the Hubble parameter versus redshift
data \cite{yi07, wan07}.

The main purpose of this work is to give out constraints on
Cardassian models with the latest observational data, including the
recently compiled 397 SNe Ia data set \cite{Hicken}, the size of
baryonic acoustic oscillation peak from the Sloan Digital Sky Survey
(SDSS) \cite{eis05}, the position of first acoustic peak of the
Cosmic Microwave Background radiation (CMB) from the five years
Wilkinson Microwave Anisotropy Probe (WMAP) \cite{komtemp}, the
x-ray gas mass fraction of clusters \cite{all04}, and the Hubble
parameter versus red shift data \cite{sim05}. As a result, we find
that the stronger constraints can be given out with this combined
data set than the former results.

This paper is organized as follows: In section 2, we give out the
basic equations of Cardassian models. In section 3, we describe the
analysis method for the observational data. In section 4, we present
the results with different data sets and some discussions for
results.

\section{THE BASIC EQUATIONS OF Cardassian MODELS}
In 2002, Freese and Lewis \cite{fre02} proposed Cardassian model as
a possible explanation for the acceleration by modifying the FRW
equation without introducing the dark energy. The basic FRW equation
can be written as
\begin{equation}
H^2=\frac{8\pi G}{3}\rho ,
\end{equation}
where G is the Newton gravitation constant and $\rho$ is the density
of summation of both matter and vacuum energy. For the Cardassian
model, which is modified by adding a term on the right side of
Eq.(1), the FRW equation is shown as below
\begin{equation}
H^2=\frac{8\pi G}{3}\rho_m+B\rho_m^n .
\end{equation}
The latter term is so called Cardassian term may show that our
observable universe as a $3+1$ dimensional brane is embedded in
extra dimensions. Here $n$ is assumed to satisfy $n<2/3$, and
$\rho_m$ only represents the matter term without considering the
radiation for simplification. The first term in Eq.(2) dominates
initially, so the equation becomes to the usual Friedmann equation
in the early history of the universe. At a red shift $ z\sim O(1)$
\cite{fre02}, the two terms on the right side of the equation become
equal, and thereafter the second term begins dominate, and drives
the universe to accelerate. If $B=0$, it becomes the usual FRW
equation, but with only the density of matter. If $n=0$, it is the
same as the cosmological constant model. By using

\begin{equation}
\rho_m=\rho_{m0}(1+z)^3=\Omega_{m0}\rho_c(1+z)^3 ,
\end{equation}
we obtain

\begin{equation}
E^2=\frac{H^2}{H_{0}^2}=\Omega_{m0}(1+z)^3+(1-\Omega_{m0})(1+z)^{3n}
,
\end{equation}
where $z$ is the red shift, $\rho_{m0}$ is the present value of
$\rho_m$ and $\rho_c=3H_0^2/8\pi G$ represents the present  critical
density of the universe. Obviously, this model predicts the same
distance-red shift relation as the quiessence with $\omega_Q=n-1$,
but with totally different intrinsic nature.

The luminosity distance of this model is
\begin{equation}
d_L=cH_0^{-1}(1+z)\int_0^zdz[\Omega_{m0}(1+z)^3+(1-\Omega_{m0})(1+z)^{3n}]^{-1/2}
,
\end{equation}
where c is the velocity of light.

The modified polytropic Cardassian universe is obtained by
introducing an additional parameter $\beta$ into the original
Cardassian model, which reduces to the original model if $\beta=1$,

\begin{equation}
H^2=H_{0}^2[\Omega_{m0}(1+z)^3+(1-\Omega_{m0})f_X(z)] ,
\end{equation}
where

\begin{equation}
f_X(z)=\frac{\Omega_{m0}}{1-\Omega_{m0}}(1+z)^3[(1+\frac{\Omega_{m0}^{-\beta}-1}{(1+z)^{3(1-n)\beta}})^{1/\beta}-1]
.
\end{equation}
Here if the   $f_X(z)$ is equal to $1$ at the same time, this model
just corresponds to $\Lambda$CDM.  The corresponding luminosity
distance of Eq. (6) is
\begin{equation}
d_L=cH_0^{-1}(1+z)\int_0^zdz[\Omega_{m0}(1+z)^3[1+\frac{\Omega_{m0}^{-\beta}-1}{(1+z)^{3(1-n)\beta}}]^{1/\beta}]^{-1/2}
.
\end{equation}

\section{DATA ANALYSIS}
For the SNe Ia data, we use the recently combined 397 data points
\cite{Hicken} , which consist with  the 307  Union data set
\cite{kow08} and 90 CFA data set. The Union set includes the
Supernova Legacy Survey \cite{astier} and the ESSENCE Survey
\cite{riess,essence,dav07}, the former observed SNe Ia data, and the
extended dataset of distant SNe Ia observed with the Hubble space
telescope. Constraints from these Sne Ia data can be obtained by
fitting the distance modulus $\mu(z)$
\begin{equation}
\mu(z)=5\log_{10}d^L+M .
\end{equation}
Here $M$ being the absolute magnitude of the object.

In 2005, Eisenstein et al.\cite{eis05} successfully found the size
of baryonic acoustic oscillation peak by using a large spectroscopic
sample of luminous red galaxy from the SDSS and obtained a parameter
$A$, which is independent of dark energy models and for a flat
universe can be expressed as
\begin{equation}
A=\frac{\sqrt{\Omega_{m0}}}{E(z_1)^{1/3}}[\frac{1}{z_1}\int_0^{z_1}\frac{dz}{E(z)}]^{2/3}
,
\end{equation}
where $z_{1}=0.35$ and the corresponding $A$ is measured to be $A =
0.469(0.96/0.98)^{0.35}\pm0.017$. Using parameter $A$ we can obtain
the constraint on Cardassian models from the SDSS.

The shift parameter $R$ of the CMB data can be used to constrain the
Cardassian models and it can be expressed as \cite{bon97}
\begin{equation}
R=\sqrt{\Omega_{m0}}\int_0^{z_r}\frac{dz}{E(z)} .
\end{equation}
Here $z_r=1089$ for a flat universe. From the five years WMAP result
\cite{spe07}, the shift parameter is constrained to be
$R=1.715\pm0.021$ \cite{komtemp}.

On the assumption that the baryon gas mass fraction in clusters is
constant which is independent of the red shift, and is related to
the $\Omega_{b}/\Omega_{m0}$, the baryon gas mass fraction can be
used to constrain cosmological parameters. Here we adopt the usually
used 26 cluster data \cite{all04} to constrain the Cardassian
models. The baryon gas mass fraction can be present as
\begin{equation}
f_{gas}^{SCDM}(z)=\frac{b\Omega_b}{(1+0.19\sqrt{h})\Omega_m}[\frac{d_A^{SCDM}(z)}{d_A^{mod}(z)}]^{1.5}
,
\end{equation}
where $b$ is a bias factor motivated by gas dynamical simulations.

Simon, Verde \& Jimenez has obtained the Hubble parameter $H(z)$ at
nine different red shifts from the differential ages of passively
evolving galaxies \cite{sim05}. The form of $H(z)$ is
\begin{equation}
H(z)=-\frac{1}{1+z}\frac{dz}{dt} .
\end{equation}
We can determine the value of $H(z)$, if $dz/dt$ is known. Recently,
the authors in \cite{gaz08} obtained $H(z=0.24)=83.2\pm 2.1$ and
$H(z=0.43)=90.3\pm 2.5$. We also add the prior $H_0 = 72 \pm 8
km/s/Mpc$ given by Freedman et al. \cite{fre01}. So now we have 11
Hubble parameter to constrain the Cardassian models.

In order to place limits on model parameters with the observation
data, we make use of the maximum likelihood method, that is, the
best fit values for these parameters can be determined by minimizing
\begin{eqnarray}
\chi^2 &=&\sum_{i=1}^{397}
\frac{[\mu(z_i)-\mu_{obs}(z_i)]^2}{\sigma_i^2}+\frac{(A-0.0469)^2}{0.017^2}+\frac{(R-1.715)^2}{0.021^2}
\\ \nonumber
&+&\sum_{j=1}^{26}\frac{[f_{gas}^{SCDM}(z_{j})-f_{gas,j}]^2}{\sigma^2_{f_{gas,j}}}+\sum_{k=1}^{11}\frac{[H(z_k)-H_{obs}(z_k)]^2}{\sigma^{2}_{H_i}}
,
\end{eqnarray}
where the $\mu_{obs},\sigma_i$ represent the corresponding
observational values for the SNe Ia, the $f_{gas}, \sigma_{f_{gas}}$
represent the corresponding observational values for the gas mass
fraction, and $H_{obs}(z), \sigma_{H}$ represent the corresponding
observational values for the Hubble parameter.

\section{RESULTS and DISCUSSIONS}
The latest observational data set, which is 397 SNe Ia + CMB + BAO +
26 gas mass fraction + 11 Hubble parameter, is used here to
constrain parameters of the original Cardassian model. By minimizing
the corresponding total $\chi^2$ in Eq. (12), we find at $1 \sigma$
confidence level $\Omega_{m0}=0.271^{+0.014}_{-0.014}$ and
$n=0.035^{+0.049}_{-0.049}$, which is shown in Fig. 1 and  is
consistent with the  $\Lambda$CDM cosmology (\cite{all04},
$\Omega_{m0}=0.25_{-0.04}^{+0.04}$). We find that combining these
observational data can tighten the constraints significantly
comparing to the results from former academic papers
\cite{gon03,wan03,sav05,yi07}. Our result gives out an even much
stronger constraint than other observational results, such as the
results from Cao 2003 with 37 SNe Ia data \cite{cao03}, Sen \& Sen
2003 with WMAP data set\cite{sen03}, Frith 2004 with about 200 SNe
Ia data set\cite{fri04}, Godlowski, Szydlowski \& Krawiec 2004 with
several different data groups\cite{god04}, Szydlowski and Czaja 2004
with SNe Ia data \cite{szy04}, Davis et al 2007 with 200 SNe Ia +
BAO + CMB data set\cite{dav07}, Zhu, Fujimoto \& He 2004
\cite{zhu04} with the dimensionless coordinate distance data of SNe
Ia + FRIIb radio galaxies + the X-ray mass fraction data of
clusters, Bento et al 2005 \cite{ben05} with SNe Ia golden sample,
Bento et al 2006 \cite{ben06} with 157 SNe Ia + BAO + CMB data set.

For the modified polytropic Cardassian model, we find   at $1\sigma$
confidence level $\Omega_{m0}=0.271^{+0.014}_{-0.015}$,
$n=-0.091^{+0.331}_{-1.908}$ and $\beta=0.824^{+0.750}_{-0.622}$.
Details for constraints are shown in Figs. 2-4, which is tighter
than that obtained in \cite{wan07}. The modified polytropic
Cardassian model reduces to the flat $\Lambda$CDM when $\beta=1,
n=0$. So the flat $\Lambda$CDM cosmology is consistent with
observations.

With these data using in this paper, we can determine when the
universe acceleration began in Cardassian models by investigating
the deceleration parameter $q(z)$. As shown in Fig. 5, we give out
the evolution of $q(z)$ in the original Cardassian expansion model,
and find the transition from deceleration to acceleration  occurs at
red shift $z_T = 0.55\pm 0.05$ in $1\sigma$ confidence level, which
is later than the result ($z_T = 0.70\pm 0.05$) obtained in
\cite{wan07}, but is consistent with the result by using the
¡®gold¡¯ sample data in \cite{bie04} ($z_T = 0.46\pm0.13$). Fig. 6
shows the evolution of $q(z)$ in the modified polytropic Cardassian
model, and we obtain the phase transition red shift is
$z_T=0.58^{+0.12}_{-0.12}$ at $1\sigma$ confidence level, which is
consistent with $z_T=0.58^{+0.17}_{-0.18}$ in \cite{wan07}. But our
result gives out a stronger constraint. With this latest data set,
we obtain very tight $1 \sigma$ error regions of the phase
transition red shift for both of the Cardarssian models, and both of
the Cardassian models' results support that the universe began to
accelerate at red shift $\sim 0.5 - 0.6$.

In addition we give the evolution of the effective equation of state
for the  modified polytropic Cardassian model. The results are shown
in Fig. 7 and from the best fit line we find the observations favor
an evolutionary quintessence dark energy model without an crossing
of $-1$ line. We also obtain that the $\Lambda CDM$ model is
consistent with the observations  and the phantom model can not be
ruled out at $1\sigma$ confidence level.

\begin{figure}
\begin{center}
\includegraphics[angle=0,scale=0.5]{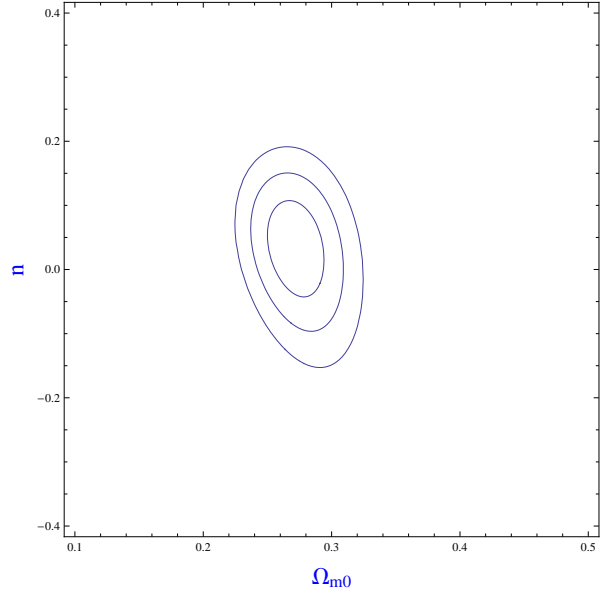}
\end{center}
\caption{Constraints on $\Omega_{m0}$ and $n$ from $1\sigma$ to
$3\sigma$ are obtained from 397 SNe Ia + CMB + BAO + 26 gas mass
fraction + 11 Hubble parameter data set for original Cardassian
model.  }
\end{figure}

\begin{figure}
\begin{center}
\includegraphics[angle=0,scale=0.5]{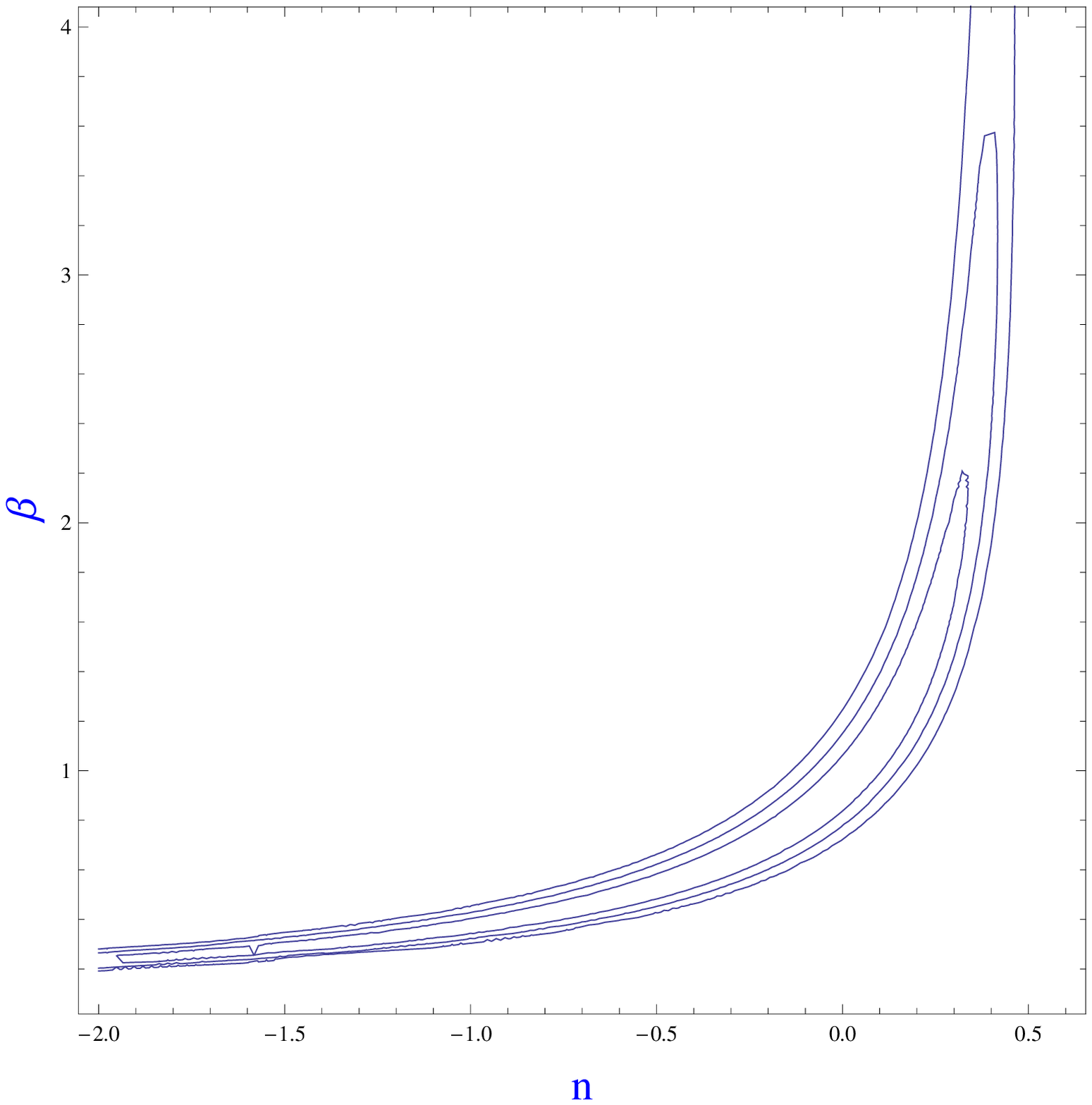}
\end{center}
\caption{Constraints on parameters $n$ and $\beta$ of the modified
polytropic Cardassian model by setting the best fit value over
$\Omega_{m0}$ from $1\sigma$ to $3\sigma$ are obtained from 397 SNe
Ia + CMB + BAO + 26 gas mass fraction + 11 Hubble parameter data
set.}
\end{figure}

\begin{figure}
\begin{center}
\includegraphics[angle=0,scale=0.5]{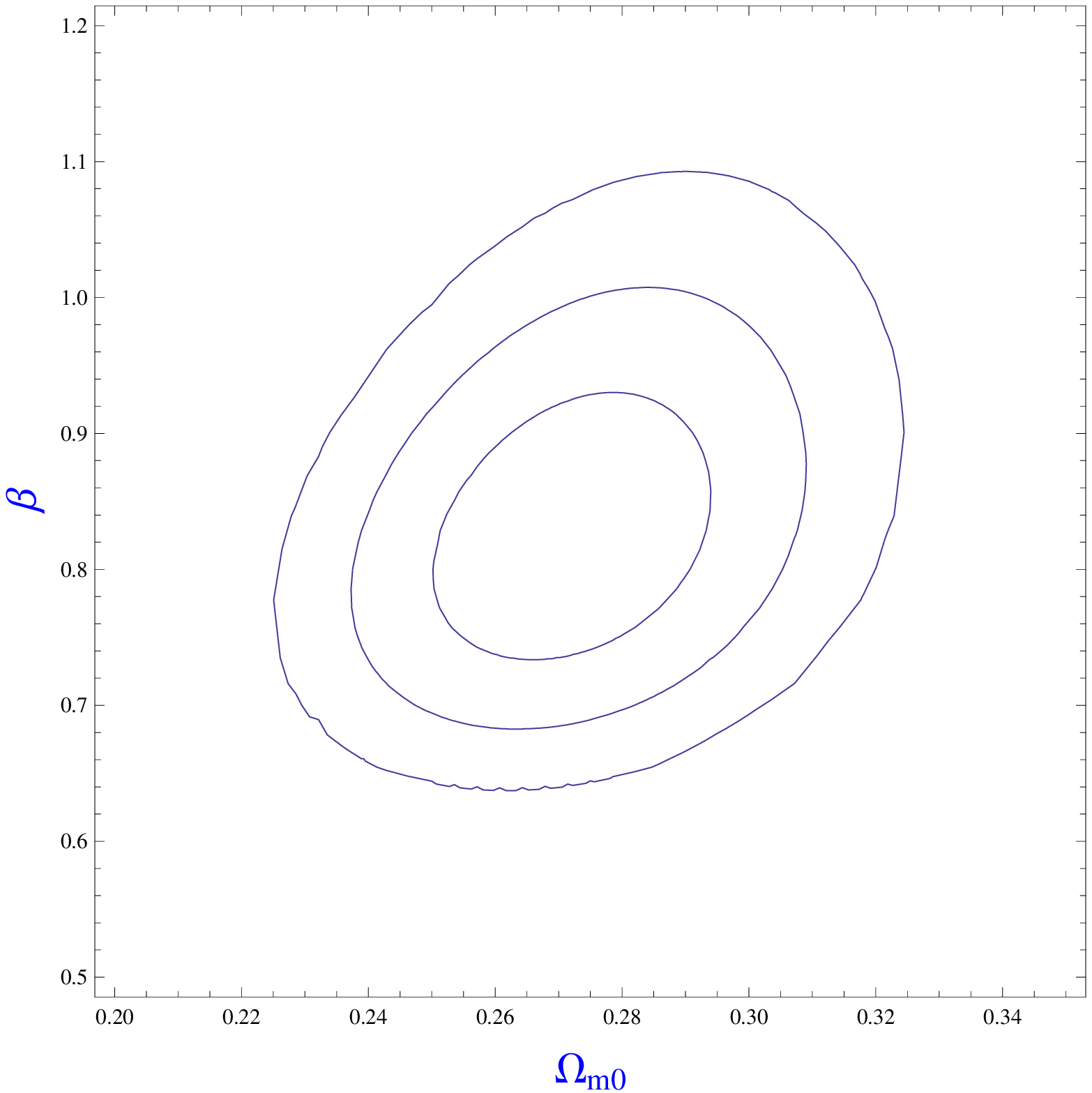}
\end{center}
\caption{Constraints on parameters $\Omega_{m0}$ and $\beta$ of the
modified polytropic Cardassian model by setting the best fit value
over $n$ from $1\sigma$ to $3\sigma$ are obtained from 397 SNe Ia +
CMB + BAO + 26 gas mass fraction + 11 Hubble parameter data set.}
\end{figure}

\begin{figure}
\begin{center}
\includegraphics[angle=0,scale=0.5]{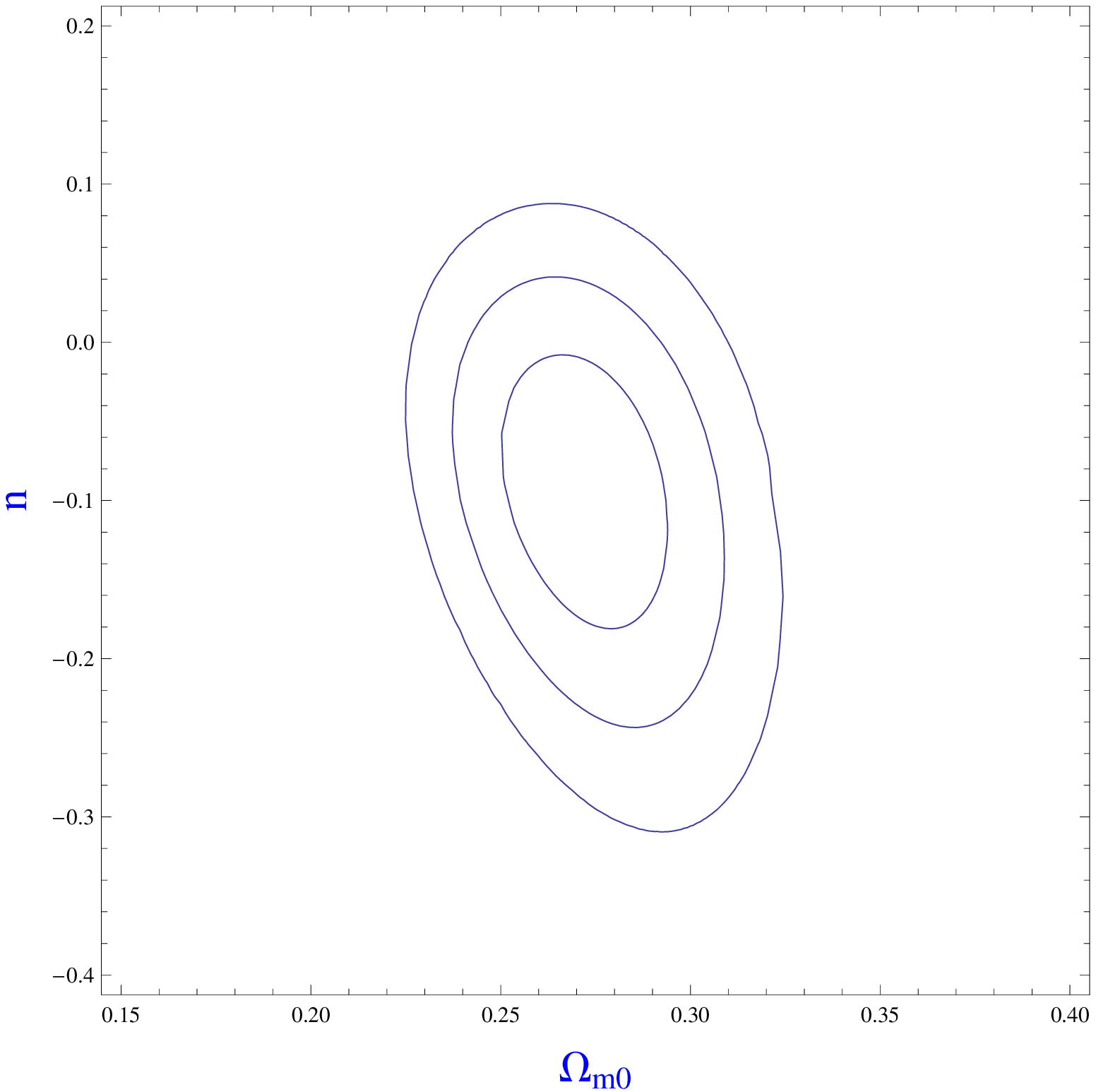}
\end{center}
\caption{Constraints on parameters  $\Omega_{m0}$ and $n$  of the
modified polytropic Cardassian model by setting the best fit value
over $\beta$ from $1\sigma$ to $3\sigma$ are obtained from 397 SNe
Ia + CMB + BAO + 26 gas mass fraction + 11 Hubble parameter data
set.}
\end{figure}

\begin{figure}
\begin{center}
\includegraphics[angle=0,scale=0.5]{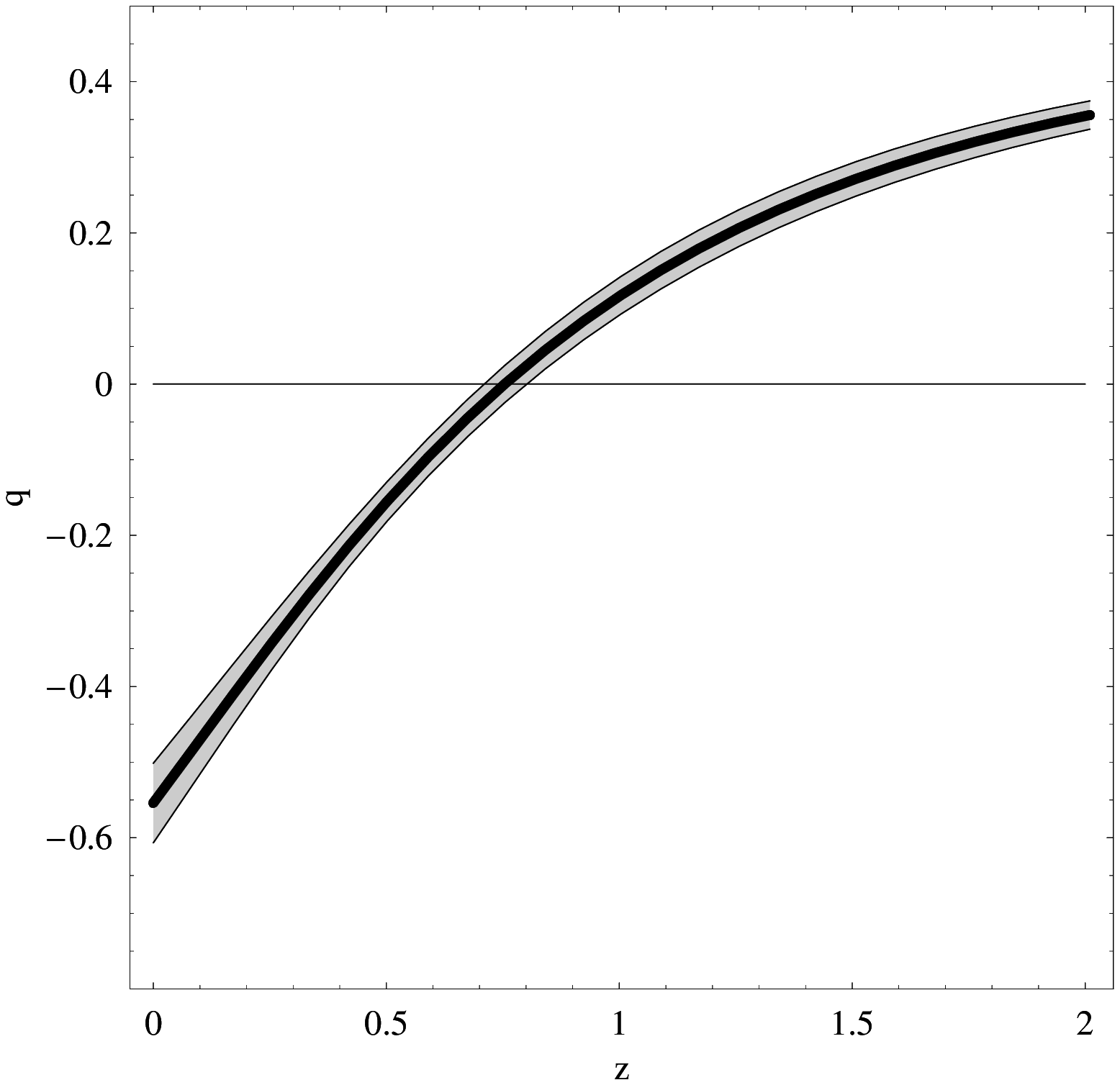}
\end{center}
\caption{The evolution of the deceleration parameter $q(z)$ for the
original Cardassian expansion model. The thick solid line is drawn
with the best fit parameters. The shaded region shows the $1\sigma$
errors.}
\end{figure}

\begin{figure}
\begin{center}
\includegraphics[angle=0,scale=0.5]{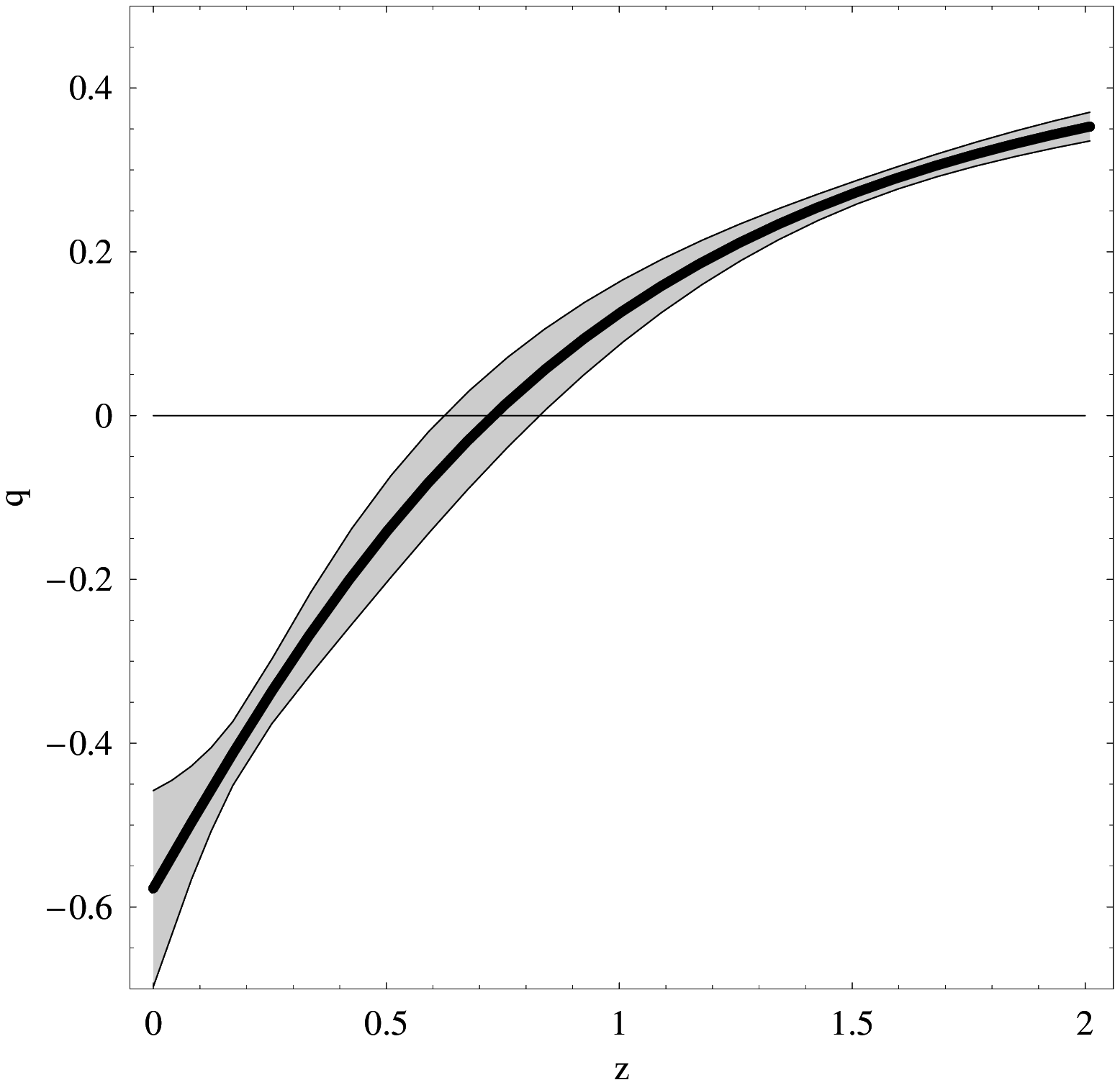}
\end{center}
\caption{The evolution of the deceleration parameter $q(z)$ for the
modified polytropic Cardassian expansion model. The thick  solid
line is drawn with the best fit parameters. The shaded region shows
the $1\sigma$ errors. }
\end{figure}

\begin{figure}
\begin{center}
\includegraphics[angle=0,scale=0.5]{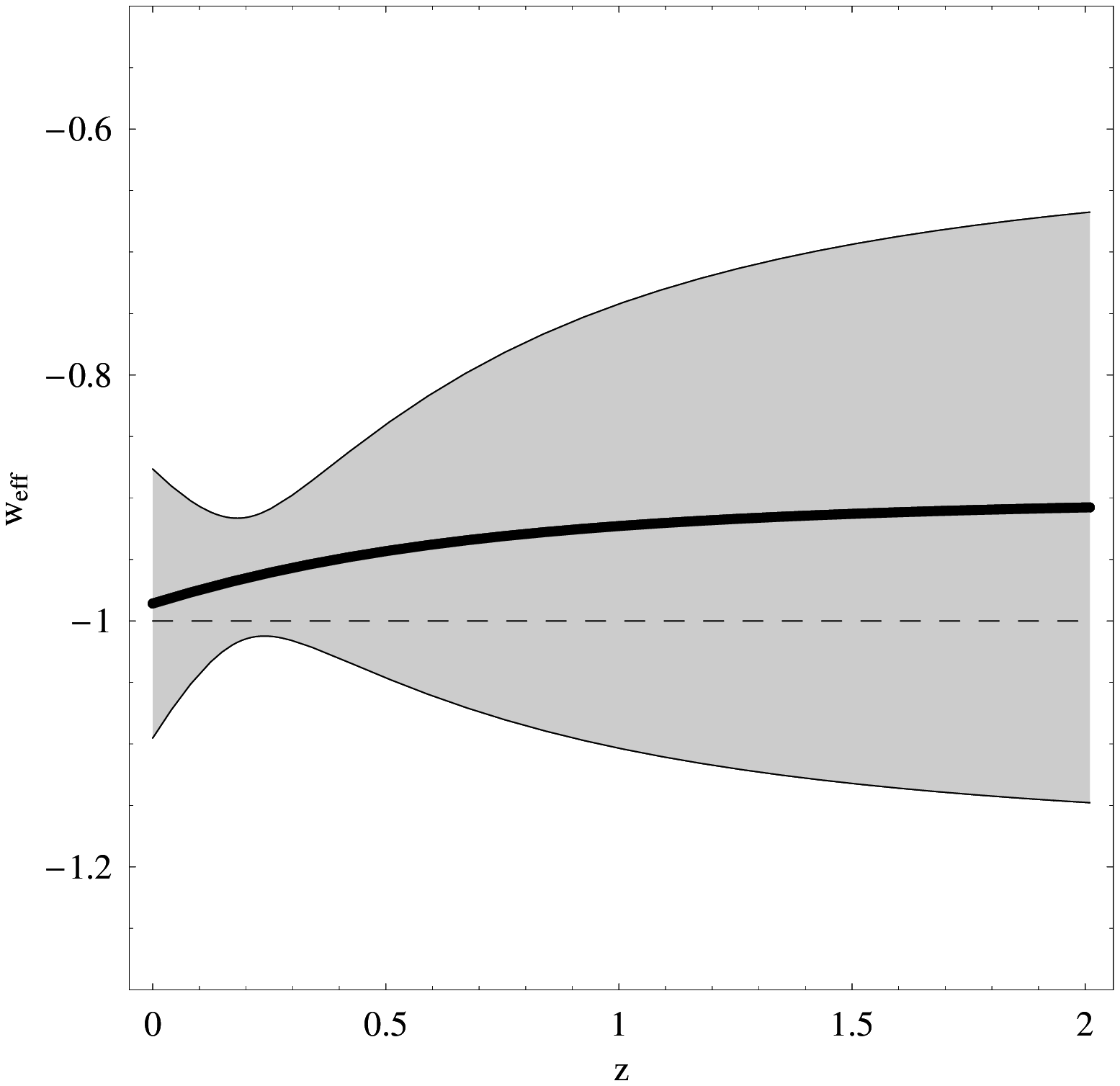}
\end{center}
\caption{The evolution of the effective equation of state for the
 modified polytropic Cardassian
expansion model. The thick solid line is drawn with the best fit
parameters. The shaded region shows the $1\sigma$ errors.  }
\end{figure}

\begin{acknowledgments}
This work is partially supported by the National Natural Science
Foundation of China  under Grant No. 10705055, the Scientific
Research Fund of Hunan Provincial Education Department, the Hunan
Provincial Natural Science Foundation of China under Grant No.
08JJ4001 and  the China Postdoctoral Science Foundation.
\end{acknowledgments}

\end{document}